\documentclass[%
 reprint,
 amsmath,amssymb,
 aps,
]{revtex4-1}

\usepackage{graphicx}
\usepackage{dcolumn}
\usepackage{bm}

\begin{document}

\preprint{}

\title{Narrow-escape times for diffusion in microdomains with a particle-surface affinity: Mean-field results}

\author{G.~Oshanin}
\email{oshanin@lptmc.jussieu.fr}
\affiliation{Laboratoire de Physique Th\'eorique de la Mati\`ere Condens\'ee (UMR CNRS 7600),
    Universit\'e Pierre et Marie Curie,
4 place Jussieu, 75252 Paris Cedex 5 France}
\altaffiliation[Also at: \,]{Laboratory J.-V. Poncelet (UMI  CNRS 2615),
       Independent University of Moscow,
       Bolshoy Vlasyevskiy Pereulok 11,
       119002 Moscow, Russia}
\author{M.~Tamm}
\email{thumm.m@gmail.com}
\affiliation{Physics Department,
    Moscow State University, Vorobyevy Gory, 119991 Moscow, Russia}
\author{O.~Vasilyev}
\email{vasilyev@fluids.mpi-stuttgart.mpg.de}
\affiliation{Max-Planck-Institut f\"ur Metallforschung,
Heisenbergstr. 3, D-70569 Stuttgart, Germany}
\altaffiliation[Also at: \,]{Institut f\"ur Theoretische und Angewandte Physik,
University of Stuttgart, D-70569 Stuttgart, Germany}

\date{\today}

\begin{abstract}
We analyze the mean time $t_{app}$ that a randomly
moving particle spends in a bounded domain (sphere) before it
escapes through a small window in the domain's boundary. A
particle is assumed to diffuse freely in the bulk until it
approaches the surface of the domain where it becomes weakly adsorbed, and then wanders
diffusively along the
boundary for a random time until it desorbs back to the bulk, and etc.
Using a mean-field approximation, we define $t_{app}$ analytically as
a function of the bulk and surface diffusion coefficients, the mean time it
spends in the bulk between two consecutive arrivals to the surface
and the mean time it wanders on the surface
within a single round of the surface diffusion.
\begin{description}
\item[PACS numbers ] 87.10.-e, 05.40.Fb, 05.40.Jc
\item[Key words ] molecular trafficking; cellular biology; diffusion; adsorption;\\
desorption; first passage times
\end{description}
\end{abstract}



\maketitle


\section{\label{sec:intro}Introduction}

A generic problem in cellular biochemistry is to estimate the time -
the so-called Narrow Escape Time (NET) -  that a randomly moving particle
spends in a bounded domain before it escapes through a small window in the domain's boundary.
A particle can be an ion, a ligand, a molecule, a protein, etc.
A confining domain can be a cell, a microvesicle, a compartment, an endosome,
a caveola, a dendritic spine, etc. A variety of processes in which the importance
of the NET problem is striking were discussed in \cite{lind,grig,schuss,benichou}.

Conventional analytical calculations of the NET rely on the
assumption that the confining surface is perfectly reflecting
everywhere, except for the escape window - an aperture of typical
size $a$. For Brownian motion, evaluation of the NET probability
density function (PDF) $F_t$ amounts to finding the solution of
the diffusion equation with mixed Dirichlet-Neumann boundary
conditions  \cite{schuss}.

In three-dimensions (3D) one finds \cite{schuss} that at sufficiently large times
the probability $S_t$ that the
particle has not reached the escape window up to time $t$ obeys
\begin{equation}
\label{classic}
S_t \sim \exp\left(- \frac{t}{t_{3D}}\right),
\end{equation}
where the symbol ``$\sim$'' signifies that one deals with the
leading in time asymptotic behavior and omits the numerical
prefactors.
The characteristic decay time $t_{3D}$ (the subscript "3D" specifies that the search for the escape window proceeds via the bulk diffusion)  in
Eq.~\ref{classic}  is given by
\begin{equation}
\label{NET}
t_{3D} = \frac{V}{4 D_0 a},
\end{equation}
where $V$ is the volume of the domain and $D_0$ is the bulk diffusion
coefficient. This result holds for any 3D bounded domain, provided
that the boundary is sufficiently smooth and the ratio $a/R$,
where $R$ is the typical size of the domain, is sufficiently
small.

The PDF $F_t$ than follows via
the relation $F_t = - dS_t/dt$. Hence, $t_{3D}$ in Eq.~\ref{NET} can be interpreted as
the mean time of the first passage to the escape window - the mean narrow-escape time.
Note that the results in Eqs.~\ref{classic}--\ref{NET} have
been obtained earlier in \cite{hill,berg,shoup} and \cite{grig}
in the special case of a sphere of radius $R$ and
the escape window being a geodesic disc
of radius $a$,
$a/R \ll 1$.

To get some idea of typical NET scales, consider
an example mentioned in \cite{lind} - search for the tubule
entrance in a vesicle
by a diffusive ligand. The vesicle size $R$ and the radius of the tubule entrance $a$ are of order of $10^{-5} \rm{cm}$ and $10^{-6} \rm{cm}$, respectively, while the ligand diffusion coefficient $D_0$ is in the range $10^{-5} - 10^{-7} \rm{cm}^2/\rm{s} $. Thus $t_{3D}$ is of order of $10^{-4}$ to $10^{-2}$ seconds, depending on the value of the bulk diffusion coefficient.
Of course, one may encounter considerably larger first passage
times for larger $R$ or smaller $a$, as well as
 under conditions of molecular crowding emerging
due to complexity of the cellular environment. In the latter case,
an effectively subdiffusive motion can emerge \cite{benichou}.
On contrary, interactions of particles with molecular motors may
induce an effective biased motion and thus reduce the NET.

The analysis based on the "perfectly reflecting wall"
assumption misses an important factor. In realistic systems, in
addition to the short-range repulsion, there are always some
attractive interactions between the surface of the domain
and the diffusive
particle. Capitalizing on ideas of Adam and Delbr\"uck, put
forward for chemoreception \cite{adam} (see also the discussion
in \cite{berg}), one may suppose that if such interactions are
sufficiently strong, the actual search for the escape window will
be a two-stage process, in which the particle will first find the
surface of the cell and then will move diffusively along the
surface until it finds the escape window. Consequently, one may
expect that in this two-stage process the rate at which the escape
window is found (and correspondingly, the apparent narrow escape
time $t_{app}$) will increase (decrease) by an amount that depends
on the surface diffusion coefficient.

\begin{figure}[ht]
\includegraphics[width=0.45\textwidth]{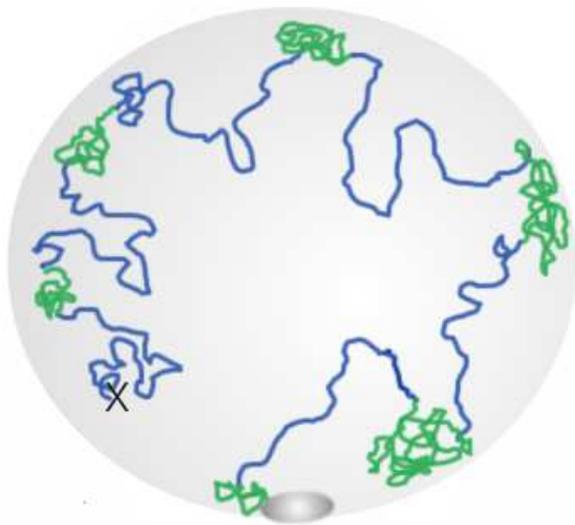}
\caption{\label{cartoon1}A path of a diffusing particle starting at point "x" and leaving the sphere through the escape window placed on the south pole. Excursions in the inner part
 of the sphere between two consecutive contacts with the surface (Brownian excursions) are marked by a blue color while the excursions along the surface - by a green color.}
\end{figure}

It would be even more realistic to
suppose that in the presence of particle-surface attractive interactions,
the search for the escape window will be an
\textit{intermittent} two-stage process  \cite{ben1,ben2,ben3,osh1,osh2}:
 a particle approaching the surface will
reversibly bind to it and, (if the barrier against lateral
diffusion is smaller than the desorption barrier), diffuse over the
surface for some (random) time $T$, after which it will desorbs back to
the inner part of the confining domain, approach it again at some
other point, reversibly bind, diffuse, et cetera. Therefore, as
depicted in Fig.~\ref{cartoon1}, a typical particle trajectory
will consist of a sequence of surface diffusion tours
followed by excursions in the inner part of the domain,
i.e., an \textit{intermittent} combination of diffusion in $2$ and
$3$ spatial dimensions.  In this case, $t_{app}$ will also acquire
a dependence on the mean  time $\tau_s$ of residence on
the surface within each round of surface diffusion. Given that
attractive interactions are always present, the result in
Eq.~\ref{NET}, based on the assumption of a perfectly reflecting
wall,
 does not provide a reliable  estimate of the actual
mean narrow-escape time $t_{app}$.

In this paper we study analytically, within a mean-field approach, the effect of the
particle-surface affinity on the mean narrow-escape times.  In
section II we define our model. In section III 
 we first rederive the result in Eq.~\ref{NET} in the spirit of
the approach discussed by Berg and Purcell \cite{berg}
in their analysis of diffusive ligands
adsorption from the extracellular space
by cell bound receptors.
Next, we
consider a special case of very strong attractive interactions,
such that a particle, once it bumps on the surface, stays there
for good and wanders along the surface until it finds the escape
window. This limit can be thought off as an analogue of the
Adam-Delbr\"uck two-stage process. Further on, we consider the
general case of an intermittent two-stage process. We develop
a mean-field approach which allows us to derive a general result
for $t_{app}$, valid for any $\tau_s$. This result represents
an interpolation formula
from which we recover  Eq.~\ref{NET} when $\tau_s \to
0$ and the result of the Adam-Delbr\"uck-type approach for
$\tau_s \to \infty$. Finally, we conclude with a
 discussion of the overall effect of the particle-surface affinity
 on the NET.

\section{Model}
\label{model}

We focus here on the simplest geometry in which the bounded domain
is a sphere of radius $R$. We base our approach on the
analysis of representative particle's trajectories, rather than on
the solution of the diffusion equation with appropriate boundary
conditions. Therefore, instead of standard settings "point
particle vs window of radius $a$", we switch here to an equivalent
formulation in which the particle has radius $a$, while the escape
window is a point on the surface. Consequently, we stipulate that
the particle covers an area $\pi a^2$, when it touches the
surface. We suppose, as well, that the particle is initially
placed at a random position on the surface of a sphere of radius
$r_0 = R - \sigma$, $\sigma = \gamma a$, where $\gamma$ is a
numerical factor of order of unity; its precise value will be
discussed below.

Diffusion coefficient of the particle in the bulk inside the
sphere is $D_0$. When the particle approaches the sphere (i.e.,
the distance between the particle and the surface of the sphere
gets smaller than $\sigma$), it becomes weakly adsorbed and starts
to diffuse, with diffusion coefficient $D_s$, along the surface of
the sphere. At every (arbitrary small) time step $\Delta t$, the
particle updates its state: with probability $p_{d}$ it may detach
from the surface and diffuse away, and with probability $1 -
p_{d}$ it stays adsorbed and continues diffusion over the surface.
The time $T$ of residence on the surface within a single surface
diffusion tour is a random variable with distribution
\begin{equation}
\label{Ps}
P_s(T) = \frac{1}{\tau_s} \exp\left(- \frac{T}{\tau_s}\right),
\end{equation}
where the mean value
\begin{equation}
\label{ts}
\tau_s = \Delta t \frac{1 - p_{d}}{p_{d}}.
\end{equation}
Our goal is to define, in this general case, the decay of the
probability $S_t$ that the particle has not found the escape window
up to time $t$, from which we will define the
tail of the first-passage distribution $F_t$ and hence,
the characteristic mean narrow-escape time $t_{app}$, as
a function of $\tau_s$, $D_0$, $D_s$, $R$ and $a$.

\section{NET problem with a particle-surface affinity}

As the first step, we rederive
the results in Eqs.~\ref{classic}--\ref{NET},
adapting to the NET problem
the approach discussed by Berg and Purcell \cite{berg} within the context of chemoreception.
Next, we will extend the developed approach over the case
when a particle has an affinity to the surface and may diffuse along the surface.

\subsection{NET problem with a perfectly reflecting wall revisited.}
\label{revision}

Let $p_{d} \equiv 1$ so that the boundary of the sphere is
perfectly reflecting. A given path of a particle during time $t$
can be then viewed as a sequence of $N$ Brownian excursions - 3D
loops connecting the points where the particle has touched the
surface; of course, $N$ is a realization-dependent random
variable. These excursions correspond to the parts of the path
marked in blue in Fig.~\ref{cartoon1}; since we focus on the case
$p_{d} \equiv 1$, the "green" parts should shrink to single
points on the sphere.

Note that the term "touching the surface", as well as the
hypothetical path depicted in Fig.~\ref{cartoon1} should be viewed
with an appropriate caution. As a matter of fact, the total number
of distinct encounters of a point particle with the surface during
a finite time interval is infinite in the limit of continuous
diffusion. To avoid this confusing behavior, one has to introduce
a finite cut-off distance of order of a realistic particle radius.

Further on, not all encounters with the surface can be considered
as independent tries in search for the escape window, but only
those Brownian excursions whose ends on the surface are separated
by a distance greater than $a$; shorter Brownian excursions should
be removed and {\it considered as} a single try. This circumstance
has been discussed in \cite{berg}. Of course, this criterion is
rather ambiguous and does not define the precise value of the
cut-off distance. In this regard, we define a Brownian excursion
as a part of a particle trajectory which starts at a distance
$\sigma = \gamma a$ away from the surface and ends up on the
surface without ever crossing it.
In doing so, we consider $\gamma$ as a fitting parameter
which will be chosen afterwards  in order to match the exact result
in Eq.~\ref{NET}.

The probability of not hitting the escape window in a single
random encounter with the surface is $1 - \pi a^2/4 \pi R^2$. If
the contacts with the surface can be taken as independent tries,
we may estimate the probability that a given path, starting at a
random location on the surface of a sphere of radius $R - \sigma$,
has not  found the escape window as a product
$\left(1 - \frac{a^2}{4 R^2}\right)^N$.
Consequently, the survival probability $S_t$ will be given by
\begin{equation}
\label{s}
S_t = \sum_{N = 0}^{\infty} P_t(N) \left(1 - \frac{a^2}{4 R^2}\right)^N,
\end{equation}
where $P_t(N)$ is the probability that the
particle "touched" the surface exactly $N$ times within time interval $t$.

Suppose now that a particle starting at $t= 0$ at distance $\sigma $ apart of the surface of the sphere
touches the surface for the first time at $t = \tau_1$, for the second time at $t = \tau_1 + \tau_2$, and etc.
Then, the probability distribution $P_t(N)$ can be defined as
\begin{equation}
\label{2} P_t(N) = E_{\tau}\left\{\theta\left(t - \sum_{k = 1}^N
\tau_k\right) \, \theta\left( \sum_{k = 1}^{N + 1} \tau_k - t
\right)\right\},
\end{equation}
where the symbol $E_{\tau}\{\ldots\}$ denotes averaging with
respect to the distribution of $\tau$-variables, while $\theta(x)$
is the Heaviside theta-function which is defined as $\theta(x) =
1$ if $x> 0$ and zero otherwise.

Using the following representation of the rectangular function:
\begin{equation}
\label{rect} \theta(t - A) \theta( B - t) = {\cal L}^{-1}
\left\{\frac{\exp(- \lambda A) - \exp(- \lambda B)}{\lambda}
\right\},
\end{equation}
${\cal L}^{-1}\left\{\ldots\right\}$ being the
 inverse Laplace transformation with respect to the parameter $\lambda$,
we perform  averaging over the distribution $P(t)$ of independent,
identically distributed $\tau$-variables and find that $P_t(N)$
obeys
\begin{equation}
\label{PP}
P_t(N) = {\cal L}^{-1} \left\{\frac{\phi_{\lambda}^N}{\lambda} \left(1 - \phi_{\lambda}\right)\right\},
\end{equation}
where
\begin{equation}
\phi_{\lambda} = \int^{\infty}_0 dt \exp( - \lambda t) P(t)
\end{equation}
is the moment-generating function of the $\tau$-variables. One may readily notice that $P_t(N)$ in Eq.~\ref{PP} is normalized, $\sum_N P_t(N) = 1$.

To evaluate $P(\tau)$, and hence, $\phi_{\lambda}$, consider the
following auxiliary problem - the survival of
 a particle, (whose initial location is uniformly distributed
 on the surface of a sphere of radius $R - \sigma$),
 which diffuses with diffusion coefficient $D_0$ within a sphere of radius $R$
 whose surface is perfectly adsorbing. The Green's function $G_{\tau}(r|r_0 = R - \sigma)$
 solution of this problem is given by \cite{carslaw}:
 \begin{eqnarray}
 \label{green}
 \displaystyle
 G_{\tau}(r|r_0) &=& \frac{1}{2 \pi R} \sum_{n = 1}^{\infty} \frac{\sin\left(\frac{\pi n r}{R}\right)}{r} \frac{\sin\left(\frac{\pi n r_0}{R}\right)}{r_0} \times \nonumber\\ 
 &\times& \exp\left(- \left(\frac{\pi n}{R}\right)^2 D_0 \tau\right).
 \end{eqnarray}
Integrating over the angular variables and $r$ we find $S_{\tau}$
- the probability that such a particle survives until time $\tau$,
from which we get the desired probability density function
$P(\tau) = - d S_{\tau}/d\tau$ that the first encounter with the
surface occurred exactly at time moment $\tau$:
\begin{eqnarray}
\label{dist}
P(\tau) &=& \frac{2 \pi D_0}{R^2  - \sigma R} \sum_{n = 1}^{\infty} n \sin\left(\frac{\pi n \sigma}{R}\right) \times \nonumber\\ &\times& \exp\left(- \left(\frac{\pi n}{R}\right)^2 D_0 \tau\right),
\end{eqnarray}
Note that the distribution in Eq.~\ref{dist} has been previously obtained in \cite{gros3} within a different context.

Before we proceed further, several remarks concerning the PDF in
Eq.~\ref{dist} are to be made. The distribution $P(\tau)$ involves
three different time scales. The smallest one corresponds to the
most probable value $\sim \sigma^2/D_0$, which means that most of
the time the particle simply bounces onto the surface almost
immediately without leaving it for any considerable distance.
Further on, at intermediate scales the distribution $P(\tau)$ has
a "fat" algebraic tail $P(\tau) \sim \tau^{-3/2}$. In this regime
$P(\tau)$ describes the probability for a random walk, commencing
at a plane bounding an infinite $3D$ system, to return back to the
plane for the first time after $\tau$ steps. As a matter of fact,
the mean $\tau$ - the mean length of Brownian excursions $\tau_b$ - is
dominated by this very regime:
\begin{equation}
\label{loop}
\tau_b = \int^{\infty}_0 d\tau \, \tau P(\tau) = \frac{R \sigma}{3 D_0} \left(1 - \frac{\sigma}{2 R}\right) \approx \frac{R \sigma}{3 D_0}
\end{equation}
and is $R/\sigma$ times larger than the most probable return time.
Finally, at times of order $R^2/D_0$, finite-size effects dominate
and the distribution $P(\tau)$ decays exponentially with time.

Consequently, the moment-generating function of a random variable $\tau$ obeys
\begin{equation}
\label{3}
\displaystyle
\phi_{\lambda} =  \frac{R}{R- \sigma} \frac{\rm{sh}\left((R - \sigma) \sqrt{\lambda/D_0}\right)}{\rm{sh}\left( R \sqrt{\lambda/D_0}\right)},
\end{equation}
from which equation we find that at sufficiently large times $t$, the distribution function $P_t(N)$ follows:
\begin{equation}
\label{dist2}
P_t(N) \sim \frac{\sqrt{5 \sigma}}{2 \sqrt{\pi R N}} \exp\left( - \frac{5 \sigma}{4 R N} \left(N - \frac{3 D_0 t}{\sigma R}\right)^2\right).
\end{equation}
This distribution is centered around the mean value $\overline{N}
= t/\tau_b$ and, at fixed $t$, decays exponentially
with $N$ on both sides of the $\overline{N}$.

Now, the asymptotic decay form of $S_t$ in Eq.~\ref{s} can be
determined in two different ways. We can either convert the sum
into an integral and use the asymptotic distribution in
Eq.~\ref{dist2}, or perform summation exactly and then invert the
Laplace transform in the asymptotic limit $t \to \infty$ ($\lambda
\to 0$). We proceed with the latter scenario. Plugging $P_{t}(N)$
given by Eq.~\ref{PP} into Eq.~\ref{s} and performing summation
over $N$, we get
\begin{equation}
\label{ss}
\displaystyle
S_t = {\cal L}^{-1} \left\{\frac{\left(1 - \phi_{\lambda}\right)}{\lambda \, \left(1 - \left(1 - \frac{a^2}{4 R^2}\right) \phi_{\lambda}\right)} \right\}.
\end{equation}
In the large-$t$ limit, the integral in Eq.~\ref{ss} is dominated by the behavior of $\phi_{\lambda}$ in the vicinity of $\lambda = 0$. Expanding
\begin{equation}
1 - \left(1 - \frac{a^2}{4 R^2}\right) \phi_{\lambda} \approx \frac{a^2}{4 R^2} - \frac{R \sigma}{3 D_0} \lambda,
\end{equation}
we find that the asymptotic behavior of $S_t$ in Eq.\ref{ss} follows
\begin{equation}
S_t \sim \exp\left(- \frac{4 \gamma R^3}{3 D_0 a} t\right).
\end{equation}
Choosing now $\gamma = \pi/4$, we see that the latter decay form
coincides with the result in Eqs.\ref{classic}--\ref{NET}. Consequently, we may interpret $t_{3D}$ in Eq.~\ref{NET} as
\begin{equation}
t_{3D} = \frac{4 R^2}{a^2} \tau_b,
\end{equation}
where the first multiplier determines the mean
number of independent tries necessary to find the location of the escape window,
while the second factor is the mean time separating independent tries - the mean length of a Brownian excursion.

\subsection{NET problem with particle-surface affinity: Adam-Delbr\"uck-type two-stage process.}

In this subsection we consider an opposite extreme case, i.e. that of
$p_{d} = 0$, so that once a particle happens to approach the
surface of the domain, it stays there for good and wanders along
the surface until it finds the escape window. In a sense,  this is
an idealized situation. Indeed, in this case the barrier against
the desorption should be infinitely large, and consequently,
the barrier against the lateral diffusion should be infinitely
large too, effectively suppressing the movement of the
particle along the surface.

Neglecting the time $\tau_b$ it will take, on average, for the particle to arrive for the
first time at some random point on the surface of the domain, we
write the probability that a particle diffusing along the surface
with diffusion coefficient $D_s$ won't find the escape window
until time $t$ as
\begin{equation}
\label{5}
S_t = \left(1 - \frac{A(t)}{4 \pi R^2}\right),
\end{equation}
where $A(t)$ is the mean area swept on the surface of a sphere of
radius $R$ by a diffusive disc of radius $a$ until time $t$ - a
two-dimensional analog a Wiener sausage. This area is defined by a
series \cite{sano}
\begin{equation}
\label{exp}
\left(1 - \frac{A(t)}{4 \pi R^2}\right) = \left(1 - \frac{a^2}{4 R^2}\right) \sum_{k = 1}^{\infty} a_k \exp\left[- \nu_k (\nu_k + 1) \frac{D_s t}{R^2}\right],
\end{equation}
where
\begin{equation}
\label{a}
a_k = \frac{1}{1 - x_0} \left(\int^1_{x_0} dx P_{\nu_k}(x)\right)^2/\int^1_{x_0} dx P_{\nu_k}^2(x),
\end{equation}
$P_{\nu}(x)$ being the Legendre functions, while $x_0 = - 1 + a^2/2 R^2$
and $\nu_k$ are the roots (numbered in the ascending order) of the
equation
\begin{equation}
\label{roots}
P_{\nu_k}(x_0) = 0.
\end{equation}
Note that the expansion in Eq.~\ref{exp} differs by a factor
$\left(1 - \frac{a^2}{4 R^2}\right)$ from the formal solution of
the trapping problem on the surface of a sphere \cite{sano}. This
difference originates from different initial conditions. Namely,
in our case a particle can be initially located at any point on
the surface (including the area covered by the trap, in which case
it disappears instantaneously, - this corresponds to finding the
escape window at a first try), while in the situation studied in
\cite{sano} the particle starts from a random point somewhere
outside the trap.

The leading asymptotic  behavior of $S_t$ in Eq.~\ref{5} is
dominated by the smallest root $\nu_1$ of Eq.~\ref{roots}. Hence,
the asymptotic behavior of $S_t$ is
\begin{equation}
\label{6}
S_t \sim \exp\left( - \frac{t}{t_{2D}}\right)
\end{equation}
with
\begin{equation}
\label{7}
t_{2D} = \frac{R^2}{\nu_1 (\nu_1 + 1) D_s},
\end{equation}
where the subscript "2D" signifies that the search for the escape window proceeds
in this case via the surface diffusion.

When $a/R \ll 1$, for the smallest root one gets $\nu_1 \approx
1/2 \ln\left(2 R/a\right)$ and, consequently, $t_{2D}$ obeys, in
the leading order  in $a/R$\cite{sano}:
\begin{equation}
\label{8}
t_{2D} \approx \frac{2 R^2}{D_s} \ln\left(\frac{2 R}{a}\right).
\end{equation}
This result has been also obtained in \cite{lind} and \cite{berg},
and earlier by Bloomfield and Prager \cite{prager}
in their calculation of the attachment rate of tail fibers to bacteriophages. Note that
$t_{2 D} \sim R^2 \ln(R)$ and thus should be much larger than
$\tau_b \sim R$, which describes the mean time necessary to reach the surface of the domain.
This means that it was quite legitimate to discard this contribution in our analysis.

\subsection{NET problem with particle-surface affinity: an intermittent two-stage process.}

We turn finally to the general case when the detachment
probability $0 < p_{d} < 1$, so that a particle, when touching the
surface, will remain weakly adsorbed and wander on the surface for
some random time $T$, then detach and diffuse in the bulk, re-attach to the surface, and etc.

Consider a path starting at a random point on the
surface of a sphere of radius $R - \pi a/4$ and suppose that this
path touched the surface of the sphere at time moment $\tau_1$,
then wandered along the surface for a random time $T_1$, detached
from the surface at time moment $\tau_1 + T_1$, subsequently
returned to the surface at time moment  $\tau_1 + T_1 + \tau_2$,
etc. Then, assuming that subsequent visits of the surface can be
considered as independent tries in search for the exit, the
probability $S_t$ that the escape window has not been found by such a
path comprising $N$ rounds of Brownian excursions followed by
subsequent surface diffusion tours, can be written as
\begin{widetext}
\begin{equation}
\label{sss}
S_t = P_t(N=0) + \sum_{N = 1}^{\infty} E_{\tau,T}\Big\{
\prod_{k=1}^N \left(1 - \frac{A(T_k)}{4 \pi R^2}\right) \, 
 \theta\left(t -  \sum_{k = 1}^N \tau_k + \sum_{k =
1}^{N-1} T_k\right) \, \theta\left(-t + \sum_{k = 1}^{N + 1}
\tau_k + \sum_{k = 1}^{N} T_k\right)\Big\},
\end{equation}
\end{widetext}
where now the symbol $E_{\tau,T}\{\ldots\}$ denotes now the averaging with
respect to both the distribution of $\tau$-variables,
Eq.~\ref{dist}, and the distribution of $T$-variables,
Eq.~\ref{Ps}.

Note that Eq.\ref{sss} tacitly assumes that
each
surface diffusion tour is an independent try in search for the
escape window, which manifests itself in the decoupling of the
average of the product into the product of average values.
This is, of course, an uncontrollable
assumption. 

On one hand, the distribution $P(d)$ of the distance $d$ between the
point where the particle detaches
from the surface and the point where it re-attaches to the surface again
after an excursion in the bulk is given by the Poisson kernel for a three-dimensional ball,
\cite{poisson}:
$P(d) \sim \frac{1}{d^3}$.
This  is a broad distribution,
such that the areas visited on the surface
in two consecutive surface diffusion tours
will not, on average, significantly overlap. 

On the other hand,
the surface of the domain
is of a finite extent and one will certainly have an oversampling
- some parts of the surface will be visited many times before the escape window is found. This will incur
some correlations in the search process, since the true survival probability $S_t$
accounts
only for the actual area swept on the surface by a particle
up to time $t$, and counts multiple visits
to the same place as a single try. In this sense, Eq. \ref{sss}
defines a lower bound on the true survival probability, precisely in the same way
as the Rosenstock (or Smoluchowski) approximation defines a rigorous
lower bound on the decay function for the trapping problem (see, e.g., \cite{ben5}).

Consequently, decoupling of correlations
defines a rigorous lower bound on the mean narrow-escape time. Given
that, as we proceed to show, $t_{app}$ obtained via such a mean-field
approach entails exact results
for $\tau_s \to 0$ and $\tau_s \to \infty$,
one may judge that it is a useful and plausible approximation.
A further discussion of this matter
goes beyond the scope of the current paper and will be addressed both analytically and numerically
elsewhere
\cite{otv2}.

Using the Laplace transform representation of the rectangular
function, Eq.~\ref{rect}, we  may conveniently rewrite
Eq.~\ref{sss} and perform straightforwardly averaging over the
distributions of $\tau$ and $T$-variables. In doing so, we get
\begin{equation}
\label{4s}
S_t = {\cal L}^{-1} \left\{\frac{1 - \phi_{\lambda}}{\lambda} +  \frac{\phi_{\lambda} \left(F_{0} - \phi_{\lambda} F_{\lambda}\right)}{\lambda \left(1 - \phi_{\lambda} F_{\lambda}\right)} \right\},
\end{equation}
where
\begin{equation}
F_{\lambda} = \frac{1}{\tau_s} \int^{\infty}_0 dT \, \left(1 - \frac{A(T)}{4 \pi R^2}\right) \, \exp\left(- \frac{T}{\tau_s} - \lambda T\right).
\end{equation}
Explicitly, $F_{\lambda}$ is given by a series
\begin{equation}
\label{f}
F_{\lambda} = \left(1- \frac{a^2}{4 R^2}\right) \sum_{k = 1}^{\infty} a_k \left(1 + \nu_k (\nu_k + 1) \frac{D_s \tau_s}{R^2} + \tau_s \lambda \right)^{-1}.
\end{equation}
Consider now the behavior of the coefficients in this series in
more detail. First, for small $a/R$ a good approximate solution of
Eq.~\ref{roots} is $\nu_k = k - 1 + 1/2 \ln\left(2 R/a\right)$
\cite{sano}, and thus the roots of Eq.~\ref{roots} grow linearly
with $k$. Second, in the leading in $a/R$ order, $a_k =
\delta_{k,1}$, where $\delta_{k,1}$ is the Kronecker-delta ($a_k =
1$ for $k=1$ and zero otherwise), correction terms to this
dependence are of order of $(a/R)^2 f_k$, where $f_k$ is a rapidly
decaying function of $k$. All this permits us, so far as we are
interested in the small-$\lambda$ (large-$t$) limit, to consider
only the first term in the series in Eq.~\ref{f} and skip the
remaining terms, giving
\begin{equation}
\label{ff}
F_{\lambda} \sim \left(1- \frac{a^2}{4 R^2}\right) \left(1 + \nu_1 (\nu_1 + 1) \frac{D_s \tau_s}{R^2} + \tau_s \lambda \right)^{-1}.
\end{equation}
Finally, notice that the first term in Eqs.~\ref{sss},~\ref{4s}
decays rapidly in the time $t$ domain, compared to the second one
(indeed, its characteristic decay time is just $\tau_b$)
and thus its contribution is negligible in the large-$t$ limit, we
get
\begin{equation}
\label{5s}
S_t \sim {\cal L}^{-1} \left\{\left(\frac{3 D_0 (a^2 + 4 \nu_1 (\nu_1 + 1) D_s \tau_s)}{12 D_0 R^2 \tau_s + \pi a R^3 } + \lambda\right)^{-1}\right\},
\end{equation}
which describes the asymptotic behavior of the survival probability in the limit $\lambda \to 0$.

\section{Results and Discussion}

Inverting Eq.~\ref{5s}, we obtain our main results:
\begin{equation}
S_t \sim \exp\left(- \frac{t}{t_{app}}\right),
\end{equation}
where
\begin{equation}
\label{app}
t_{app} = \frac{12 D_0 R^2 \tau_s + \pi a R^3 }{3 D_0 (a^2 + 4 \nu_1 (\nu_1 + 1) D_s \tau_s)}
\end{equation}
is the mean NET for diffusion in a sphere with particle-surface affinity. Note that for $\tau_s \to 0$
we recover the exact result in Eq.~\ref{NET}, while for  $\tau_s \to 0$ we find from
Eq.~\ref{app} the exact result in Eq.~\ref{7},
specific for the Adam-Delbr\"uck-type two-stage search process.
We recall, as well, that Eq.~\ref{app} defines a rigorous lower bound on the actual mean narrow-escape time for any $\tau_s$.

It is expedient to cast the result in Eq.~\ref{app} into a
physically meaningful form, using the characteristic times
introduced in Eqs.~\ref{ts},~\ref{loop},~\ref{7}:
\begin{eqnarray}
\label{final}
t_{app} &=& \frac{\tau_b + \tau_s}{\frac{a^2}{4 R^2} + \frac{\tau_s}{t_{2 D}}}.
\end{eqnarray}
This equation has a transparent physical
meaning: the numerator on the right-hand-side of Eq.~\ref{final} defines
the overall time spent, on average, in a single Brownian excursion in the bulk followed by a surface diffusion tour, while the denominator defines the average fraction of the sphere surface covered
within a single tour of surface diffusion. Therefore, $t_{app}$ equals the time consumed by a Brownian excursion
and a single surface diffusion tour, times the number of tries necessary to cover the whole
surface. We note that a similar result has been obtained in \cite{bruinsma,gros2} for search by a diffusive protein for a specific binding site on a DNA molecule.

Further on, we rewrite Eq.~\ref{final} formally as
\begin{equation}
t_{app} = t_{3D} \left(1 + \frac{\tau_s}{\tau_b}\right)/\left(1 + \frac{4 R^2}{a^2} \frac{\tau_s}{t_{2D}}\right).
\end{equation}
One notices next that $t_{app}$ is a monotonically increasing (decreasing) function of $\tau_s$ if $t_{2 D} > t_{3 D}$ ($t_{2 D} < t_{3 D}$). Consequently, $t_{app} > t_{3 D}$ when
\begin{equation}
\label{ineq}
\frac{D_0}{D_s} > \frac{\pi}{3} \nu_1 (\nu_1 + 1) \frac{R}{a}.
\end{equation}
Typically, $D_s$ is less by two or three orders of magnitude than $D_0$ (may be even less under the conditions of molecular crowding \cite{ben6}), which means that the ratio on the left-hand-side of the inequality in Eq.~\ref{ineq} is of order of $10^2$ to $10^3$. For the example, mentioned in the Introduction, - search for the tubule
entrance in a vesicle
by a diffusive ligand, one has $R/a \sim 10^1$ so that the right-hand-side of Eq.~\ref{ineq} is of order of unity
and
the inequality in
Eq.~\ref{ineq} evidently fulfills.
This means that for this example the particle-surface affinity will generally lead to \textit{larger}  mean narrow-escape times, compared to the estimate based on the assumption of a perfectly reflecting wall. To inverse the inequality, one will need the ratio $R/a$ to be or order of $10^3$ to $10^4$, which may be realized, say, for catalytic reactions in microporous media ($R$ being the radius of a pore and $a$ - radius of a catalytic site). In this case, indeed, one may expect that particle-surface affinity will reduce the effective times of the first-passage to the catalytic site and thus enhance the reaction rate.

\section{Acknowledgments}

G.O. is partially supported by Agence Nationale de la Recherche
(ANR) under grant ``DYOPTRI - Dynamique et Optimisation des Processus de
Transport Intermittents''.

\end{document}